# PRODUCTIVITY DEVELOPMENT IN THE CONSTRUCTION INDUSTRY AND HUMAN CAPITAL: A LITERATURE REVIEW

Matthias Bahr and Leif Laszig

Department of Civil Engineering, HS Biberach, Germany

*ABSTRACT*

*Tis paper is a literature review focusing on human capital, skills of employees, demographic change, management, training and their impact on productivity growth.*

*Intrafirm behaviour has been recognized as a potentially important driver for productivity. Results from surveys show that management practices have become more structured, in the sense of involving more data collection and analysis. Furthermore, a strong positive correlation between the measured management quality and firm performance can be observed. Studies suggest that there is a positive association between management score and productivity growth.*

*The lack or low level of employees' skills and qualifications might be in different ways a possible explanation for the observed slowdown of productivity growth. The main reason for the decline in skilled labor is the demographic change. Construction sectors are increasingly affected by demographic developments. Labour reserves in construction are largely exhausted. Shortage of qualified workforce is impacting project cost, schedules and quality.*

## 1. INTRODUCTION

This paper discusses the object of human capital with regards to construction sectors in the form of qualitative (management practices/quality) and quantitative (shortage of workforce) aspects and their impact on productivity (by the example of USA and Germany).

A downward trend on the productivity growth rate for the construction sectors of western industrial countries has been observed since the 1970s of the twentieth century. The phenomenon of decreasing productivity growth rates is known as productivity slowdown. It describes both the phases of positive but decreasing growth and also of negative growth. Being a good indicator of countries' competitiveness and industry location attractiveness, pinpointing the determinants of productivity has become increasingly interesting to economists and policy makers.

Noted general causes for the above described productivity development in the national construction industries can be mainly seen in low investment (in capital equipment), the unique fragmentation and heterogeneity (e.g. firm size) of construction sectors, customer-driven production, slow diffusion and adaption of innovation, increasing project complexity, the structural change and demographic developments and their impact on human capital (e.g. in the sense of management quality and workforce shortage or skill mismatch).

                                                                                                 1

On the other hand, the recent economics literature has pointed out the existence of substantial differences in productivity levels, even amongst similarly developed countries such as eg the US and Germany. The OECD for example shows that aggregated labor productivity growth in Germany has lagged substantially behind the US for the last two decades: In 2015 labor productivity for the US total industry amounted to 60,38 USD/hour and for the German total industry to 42,49 EURO/hour, while labor productivity for US construction amounted to 37,18 USD/hour and for German construction to 25,80 EURO/hour. [1]

Intrafirm behavior has long been recognized as a potentially important driver for productivity. Results from surveys [2, 3, 4] show that management practices have become more structured, in the sense of involving more data collection and analysis. Furthermore, a strong positive correlation between the measured management quality and firm performance can be observed. Studies suggest a positive association between management score and productivity. This score is a measure of how structured management is at the establishment level and may as such be interpreted as a measure of management "quality". According to *Broszeit et al.* [4] the quality of management, measured by this management score, has increased among German establishments between 2008 and 2013, but still lags behind a comparable measure for the US.

National construction sectors of Western industrial countries are increasingly affected by demographic developments. The main reason for the decline in skilled labor is the demographic change. This includes the low birth rate of the past years and decades. The baby boomers of the 1950s and 1960s approach the age of retirement and retire from working life. The succeeding generations entering the employment and training market are less compared to the retiring employees and thus, the supply of skilled manpower is receding further with occurring shortages. Labor reserves in construction have thus been largely exhausted. Age-equitable and age-appropriate workplaces are needed, not only to reduce the incidences of early retirement. The risk that qualified staff moves to other sectors seems to be a further concern. Enterprises seem to fill the increasing skilled workers' gap through the assignment of foreign workers (from the European Community). According to the OECD the lack or low level of employees' skills and qualifications (e.g. in the form of quantitative skill shortage or qualitative skill mismatch) might be in different ways (e.g. in the use of digital technologies) a possible explanation for the observed slowdown in the growth of productivity.

In the short term, construction labor shortages are affecting construction compensation, training and operations. Firms are boosting pay, adding new benefits and taking other steps to make construction careers even more rewarding. They are launching new training programs and expanding existing ones. They are adopting new technologies and new techniques to increase productivity.

The longer-term consequences of these labor shortages are likely to be more significant. As the cost of labor goes up and firms look at ways to become more efficient, technology is likely to play an even greater role in the construction process. This could lead to significant changes in the way projects are planned and executed and it could also alter the nature of much of the work being performed by the sector's employees.

## 2. DEVELOPMENT OF PRODUCTIVITY GROWTH

The past 200 years have been an era of extreme productivity increase unique in the history of mankind. The techno-scientific revolution (second industrial revolution) has led to annual productivity growth of two to three percent for the US industry. The bell-shaped curve in Figure 1 starts after 1800 and is again approaching zero - as to when this will occur is subject to controversy. [5]

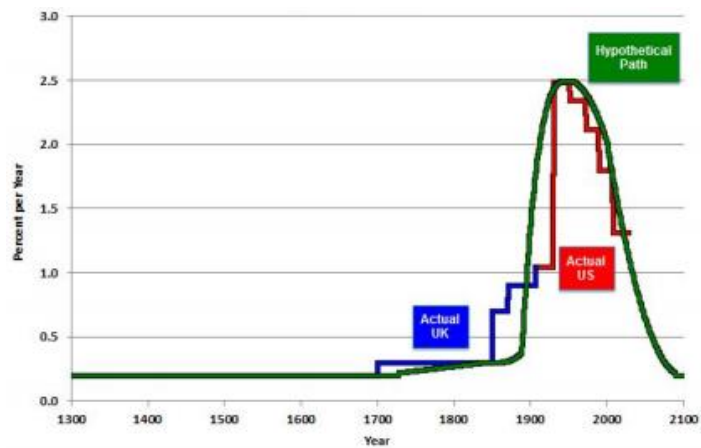

Figure 1. Productivity growth (in GDP per capita) 1300 - 2100
actual and hypothetical (forward) projection [5]

The growth of productivity (eg measured as output per hour) slowed markedly after 1970. While puzzling at the time, it seems increasingly clear that the one-time-only benefits of the Great Inventions and their spin-offs had occurred and could not happen again. Diminishing returns set in, and eventually all of the subsidiary and complementary developments following from the Great Inventions of the Second Industrial Revolution had happened. All that remained after 1970 were second-round improvements. Though *Gordon's* findings apply to the US these findings are also valid for other industrialized countries of the western world.

*Gordon* [5] identifies for the US economy possible reasons ("headwinds") that will limit future potential growth and hold it below the pace which innovation would otherwise make possible. These "headwinds" include the (1) end of the "demographic dividend" which is now in reverse motion; a (2) rising inequality since the growth in median real income has been substantially slower than all of these growth rates of average per-capita income discussed thus far; the (3) effects of the interaction between globalization and ICT (e.g. foreign inexpensive labor competes with American labor not just through outsourcing, but also through imports); a (4) twin educational problems of cost inflation in higher education and poor secondary student performance (the US is steadily slipping down the international league tables in the percentage of its population of a given age which has completed higher education); the (5) consequences of environmental regulations and taxes that will make growth harder to achieve than a century ago (since today's rich nations of North America, Europe, and Japan were not regulated in the same way during their 20th century period of high growth); and the (6) overhang of consumer and government debt.

How quickly the human history of productivity impetus will end depends also on the possibilities and speed of additional automation in the fields of production and services. At present, about half of the automation potential in materials production in the highly developed industrial countries has been exhausted (Figure 2). Until ninety percent of materials production and the desirable part of services have been worldwide automated, another hundred years may pass. [6]

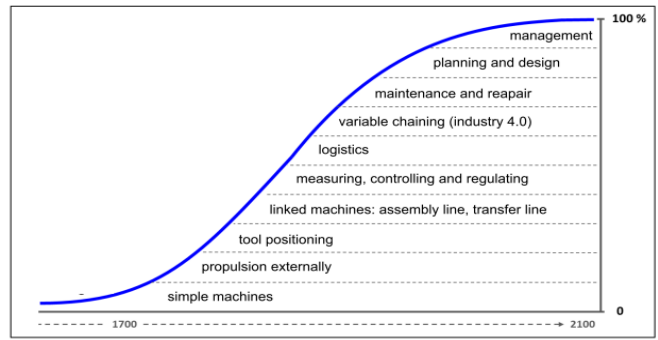

Figure 2. Grade and potential of automation [6]

The significance of automation plays an ever more important role in the global economy. It is one of the drivers of our economy and simplifies many aspects of everyday life. While an infusion of new labor seems unlikely automation is emerging as a possibility.

Automating construction is especially challenging because structures must be tailored to sites and tastes in ways that do not lend themselves to mass production. The ability to customize and transition quickly between short production runs is key. Successful approaches in the construction industry to install automation technology has thus far been limited. Civil engineering is mainly characterized by one-of-a-kind buildings as well as changing work sites with generally unfamiliar working conditions (eg weather, soil condition, logistics). Such boundary conditions make the economical application of automation (and digitalization) difficult. In contrast to classic industrial manufacturing, a construction site is significantly less well suited to full automation because of the heterogeneous tasks required and the changing conditions. That is the reason that construction jobs are carried out then as now mostly by hand and accompanied by hard physical effort or with simple technical support. However, the most recent developments in construction equipment do make increasingly semi-automated manufacturing processes possible. [7] In order to counteract the declining or in some cases, stagnating productivity trend, some hopes rest upon automation (and digitalization) and the increasing deployment of ICT that implies. Already today, many parties at construction sites benefit from the introduction of digital technologies.

Thus, a way of raising the construction industry's capacity is through technological innovation. From a technical point of view, construction is regarded as insufficiently innovative. Productivity increasing innovations in construction differ from other industry sectors and service sectors. They are strongly process-oriented, incremental (to happen stepwise) and in many cases they are related to specific problems occurring suddenly and one-time (without sustainable treatment). The attempt is often made to apply measures of rationalization and productivity advances to the construction industry that work for stationary industries. However, because of the basic differences between these businesses and the peculiarities in the "manufacture" of buildings and infrastructure, these measures cannot be transferred directly to the construction sector.

The *McKinsey Global Institute* estimates in an anecdotic essay that the worldwide lost economic potential due to the low productivity in construction, has reached an annual level of about 1.6 billion USD. This corresponds to about two percent of annual economic output globally. On the national level, higher productivity would also represent an advantage in international competition. [8]

In order to compare and analyze productivity developments on a sectoral level, labor productivity (usually measured as output per hour) and Total Factor Productivity (TFP) and in some cases as well additional factors (key indicators) are being increasingly studied. As our data source for this, we draw upon the EU KLEMS database [1]. It consists of harmonized data from the national federal statistical offices and is therefore internationally compatible.

The EU KLEMS database was most recently updated at the end of 2017 (with a revision in July 2018). The data for construction are accrued in accordance with ISIC Rev. 4; Code F and lie mostly in the period of 1980 to 2015. The advantage of the EU KLEMS data lies in the fact that it allows a discussion of the long-term development trends in productivity and especially also for the TFP. The TFP as the central means of productivity measurement expresses the extent to which the growth of value creation takes place independent of the change in factor input. It is therefore not caused by a rise or fall in the deployment of such production factors as labor and capital. The cause for this part of the growth can primarily be ascribed to technical advances. Thus, the sector related TFP represents an indicator, not without controversy among growth researchers, that expresses how much a sector in the respective national economy contributes to technical advance. [9, 10]

In an international comparison of large economies (of the European Union) and the US, the development productivity growth for the national construction sectors differs considerably. Labor productivity in the US construction industry has performed significantly below average in comparison with other economies considered here (see Table 1).

Table 1. Development of labour productivity – EU countries and USA 2000 to 2015 [1]

| nation (sector) | 2000 | 2005 | 2010 | 2015 | 2000 - 2005 | 2005 - 2010 | 2010 - 2015 | 2000 - 2015 |
|---|---|---|---|---|---|---|---|---|
| | index / absolute value (2010 = 100) | | | | growth rate in percent p.a. | | | |
| France (construction) | 118 | 115 | 100 | 93 | 0.12 | -2.54 | -1.33 | -1.26 |
| Germany (construction) | 100 | 102 | 100 | 101 | 0.63 | -0.45 | 1.00 | 0.13 |
| UK (construction) | 95 | 100 | 100 | 102 | 0.73 | -1.43 | 2.39 | 0.24 |
| USA (construction) | 113 | 105 | 100 | 96 | -1.20 | -1.55 | -0.43 | -1.07 |
| Germany (total industries) | 89 | 97 | 100 | 104 | 1.79 | 0.80 | 1.12 | 1.14 |
| USA (total industries) | 84 | 93 | 100 | 99 | 2.13 | 1.42 | 0.18 | 1.17 |

The following Figure 3 demonstrates three important findings for the development of labor productivity: (1) the annual growth rate of labor productivity in construction differs significantly across the observed countries. (2) Further there seems to be no general productivity trend in construction, all countries assessed are equally affected by. (3) The productivity development in the period prior to the global economic crisis (1997 to 2008) appears less volatile compared to the period from 2009 to 2015.

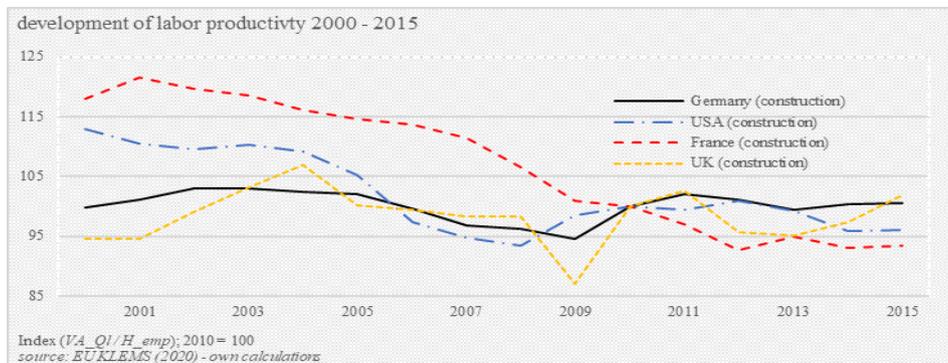

Figure 3. Development of labor productivity – EU countries and USA 2000 to 2015 [1]

The average TFP growth, which is commonly used as a measure for general technical progress, is negative for all considered countries except for the United Kingdom / UK (see Table 2). The average real growth rate of TFP in the US construction industry in the period from 2000 to 2015 amounts to -1.67 percent, again well below the average for the economy as a whole (+0.46 percent) and also well below-average in the international comparison among construction sectors. This decline in TFP in construction, although already shown in previous studies, is surprising. A possible explanation for the negative TFP growth would be, on the one hand, the fact that this measure has already been adjusted for the technical progress embodied in new investment goods. On the other hand, because the TFP is generally calculated as a residual, any measurement errors in the national accounts, especially in the construction industry (eg because of difficult measurement of the number of hours worked), could play a certain role. [11]

Table 2. Development of TFP (*TFPva_l*) – EU countries and USA from 2000 to 2015 [1]

| nation (sector) | 2000 | 2005 | 2010 | 2015 | 2000 - 2005 | 2005 - 2010 | 2010 - 2015 | 2000 - 2015 |
|---|---|---|---|---|---|---|---|---|
| | index / absolute value (2010 = 100) | | | | growth rate in percent p.a. | | | |
| France (construction) | 119 | 115 | 100 | 91 | 0.20 | -2.60 | -2.13 | -1.47 |
| Germany (construction) | 102 | 100 | 100 | 99 | -0.06 | -0.24 | 0.69 | -0.13 |
| UK (construction) | 96 | 102 | 100 | 100 | 0.85 | -0.92 | 1.99 | 0.03 |
| USA (construction) | 126 | 116 | 100 | 98 | -1.55 | -3.08 | -0.27 | -1.67 |
| Germany (total industries) | 97 | 99 | 100 | 103 | 0.57 | 0.38 | 0.90 | 0.47 |
| USA (total industries) | 95 | 99 | 100 | 101 | 0.94 | 0.35 | 0.29 | 0.46 |

1: 2014 (2015 not available)

The two sections are addressing the productivity affecting aspects of management practices / quality (3.1) and shortage of workforce (3.2). Other productivity affecting aspects relative to construction (as described above), though also crucial, are not taken into account at this point.

## 3. HUMAN CAPITAL AND PRODUCTIVITY

### 3.1. Management Practices/Quality and Productivity

"If management is related to firm level productivity, then differences in management scores across countries may be able to explain productivity differences across countries as well." The underlying assumption concerning this relationship is that management positively affects firm performance through several channels [2]:

- First, the management practices (that they inquire about in the MOPS survey) show a certain level of structure in the firm, which make production and problem-solving processes more efficient and thereby increases productivity.

- Second, a higher level of employee supervision may lead to more pressure transferred to the employees, but also to a higher motivation level, employee effort and job satisfaction. [12] This in turn increases productivity as well. [13]

- Third, there is a self-sorting process of workers, resulting from the fact that workers who are less productive leave the company or are not even hired. [14] This is in line with *Bender et al.* [15], who find that better-managed firms have a higher share of workers and managers with above-average human capital than less-well managed firms.

In this context, intrafirm behavior has long been recognized as a potentially important driver of productivity. Based on the World Management (WMS) Survey questionnaire, in 2010 the US Census Bureau carried out the Management and Organizational Practices Survey (MOPS). The data include information on over 30,000 manufacturing firms in the US and provide information on management practices and firm characteristics for the years 2005 and 2010. Results from this data show that management practices have become more structured, in the sense of involving more data collection and analysis (eg for production targets or bonus payments). Furthermore, a strong positive correlation between the measured management quality and firm performance was observed. [16]

*Broszeit et al.* [4] built on this research and conducted a similar survey among establishments in Germany, the German Management and Organizational Practices Survey (GMOP). *Broszeit et al.* collected information on over 1,900 establishments across German manufacturing industries for the years 2008 and 2013. They show that there is substantial heterogeneity in the management score across establishments. This indicates widespread differences in management practices within Germany. *Broszeit et al.* attempt to explain the observed heterogeneity in management quality by using observable firm characteristics relating to earlier work by Bloom and van Reenen [2, 3]. Furthermore, they investigate the link between management and labor productivity and find that the management score is positively and robustly related to labor productivity. Given that the calculated management scores for Germany are, on average, lower than in the US, lower management quality may explain at least partly the productivity differences between the US and Germany that were alluded to above. It is conceivable that the relatively lower labor market flexibility in Germany prevents or hinders the use of some management practices concerning human resources (eg hiring and firing, promotion or bonuses). Additionally, higher levels of collective bargaining, union coverage and works councils may have similar dampening effects.

Further, the analyses of *Broszeit et al.* show that establishments with high management scores are generally larger, have higher shares of managers with university degrees, are more likely to be foreign-owned, to be active abroad and have a works council. They also appear to be more productive by means of labor productivity. In other words: the larger the firm, the more structured and, in this sense, "better" is the management on average. Medium-sized establishments are thus doing better than small establishments, but on average lag behind large establishments in terms of their management structure.

Given the comparatively low level of management scores for the SME's (which is the typical business form in construction), there is substantial potential for catching up. Improving management practices among this group of establishments could lead to gains in productivity, even if these may be relatively lower than those reaped by large establishments. This apparent underperformance of small and medium sized firms may also be part of an explanation for the productivity differences observed between Germany and the US. It also links to a broader international debate on growing productivity dispersion.

Future research might therefore explore to what extent management practices, as a form of tacit knowledge of the production process, diffuse too slowly among construction firms and whether complementary investment, eg computerized information, can help mitigate this process. [17]

### 3.2. Workforce Shortage and Productivity

According to the *OECD* [18] the lack or low level of employees' skills and qualifications (e.g. in the form of quantitative labor shortage or qualitative skill mismatch) might be in different ways a possible explanation for the slowdown in the growth of productivity.

Construction sectors are increasingly affected by demographic developments (decrease of birth rate, aging population, increase of migration). Shifts in workforce age demographics are normal as one generation begins to age out and the next generation enters. However, current US statistics show that the construction industry appears to be aging faster than it can replace older workers. The contraction of the construction industry and its accelerated aging since 2005 appear to be driven by two trends. First is the Baby-Boomers' progression through the lifecycle, which is not unique to the construction industry. Second – and distinct from other industries – the reduction in the industry's employment of younger workers. The decreasing representation of younger workers in the industry indicates a failure to maintain sufficiently-sized replacement cohorts.

Various studies of the past and also recent statistics provide empirical evidence that the construction sector is increasingly experiencing bottlenecks in the provision of qualified employees. In a survey of thirty leading enterprises in the German construction industry by *PricewaterhouseCoopers* [19], the surveyed companies rated the current and also the future bottleneck of skilled labor (wage and salaried employees) as the biggest challenge, followed a far behind by factors such as labor cost, material and energy prices or economic policy conditions. Particularly small and medium-sized establishments at locations not situated in major cities (congested areas) fight for recruitment of capable staff. A similar result was reached in a study dated 2014 (throughout Germany, 223 companies were surveyed). Approximately 21 percent replied that they are confronted with a pronounced shortage of skilled personnel and 55 percent with phased bottlenecks in the provision of staff. Thus, the provision of junior staff constitutes a significant condition for construction companies.

The following paragraphs discuss the main reasons for the observed labor shortage (age structure and aging workforce, employment versus unemployment, shift of educational level, withdrawal or exit from the sector) and the necessary implementation of counter measures.

With regard to the German construction industry, between 2008 and 2018 the proportion of older waged employees has further increased: about 22 percent of the waged employees group were already 55+ aged in 2018. Compared to the year 2008 this means an increase of almost 60 percent. In contrast to this, especially the share of middle-aged employees (25 to 44 years old) decreased (from 50 percent to about 43 percent). This situation has relaxed compared to the previous year, since the employment numbers for the younger age cohorts (15 to 34 years) were once again slightly raised (Figure 4).

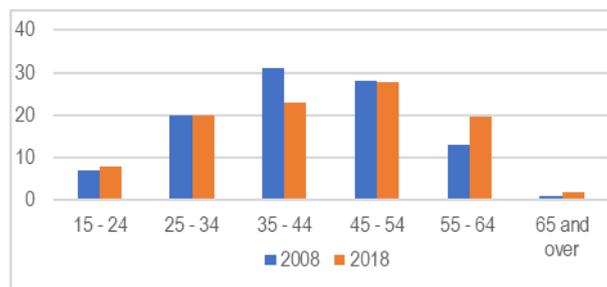

Figure 4. Age structure of industrial workers in the German construction industry (percent) [20]

In the past years, the age structure of the German salaried employees in construction followed a similar trend as that for the waged employees. Looking back over the last ten years the aging process for the salaried employees is progressing at a slower rate compared to waged employees, however at higher age. In absolute terms, almost 150,000 salaried and waged employees need to be replaced for reasons of retirement over the coming ten years. This means that the safeguarding of human resources will be an overriding issue in the coming years for the German construction industry. [20]

The latest US labor force statistics from the 2018 Current Population Survey show that the construction industry continues to struggle to attract younger workers. While workers under the age of 25 comprised 12 percent of the US labor force, their share in the construction industry reached only 9 percent in 2018. Meanwhile, the share of older construction workers ages 55+ increased from less than 17 percent in 2011 to almost 22 percent in 2018.

Figure 5 shows that, as of 2018, only about 9 percent of construction workers were 16 to 24 years old, less than the employment share of this age group in all industries. Around 69 percent of construction workforce were in the prime working years of 25 to 54, compared to 64 percent in overall workforce. The share of workers ages 55 and older was 22 percent in construction, implying that a substantial portion of workforce would retire in near future. The relative greater share of workers in construction in the 35 to 55 age group, mostly Gen X-ers, reveals the current challenge. Gen X is a smaller generational group that the Baby Boomers were.

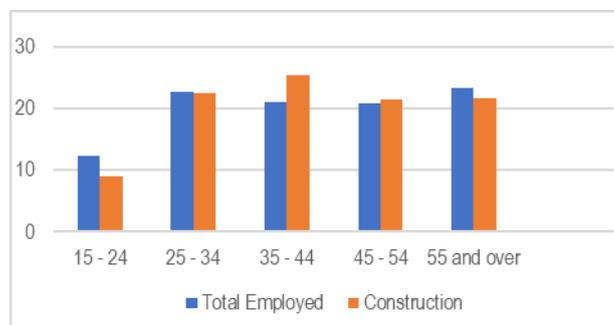

Figure 5. Age structure US Construction Industry vs All Industries (percent) [21]

Analysis of the age distribution of US construction workers over time reveals that the US construction workforce is aging, with the share of older workers ages 55+ rising from 17 percent in 2010 to almost 22 percent in 2018. At the same time, the proportion of workers ages 25 to 54 declined from 75 percent to 70 percent. This change in age composition of construction labor force is largely because the last elements of the Baby Boomer generation are entering the 55+ age group. The share of younger construction workers ages 19 to 24 edged up to 9.0 percent from 8.6 percent.

The explanations above show that, labour reserves in the construction industries have been largely exhausted: In Germany, the number of unemployed skilled construction workers reached an historical low point in 2018. Further, the numbers of blue-collar apprentices and younger people coming through the system were not sufficient to replace those who leave for retirement in the past years the companies succeeded to further expand their employment levels. Enterprises seem to fill the increasing skilled workers' gap through the assignment of foreign workers from the community of the European Union. Following this, the employment share of foreign workers (relating to the main construction sector) rose from around 8 percent in 2009 to 18 percent in 2018. [22]

*AGC* of America and its partner *Autodesk* surveyed construction firms during the summer of 2019 to evaluate the extent and impacts of workforce shortages in the industry. [23] The survey found that an overwhelming majority of the nearly 2,000 construction firms that responded are having a hard time finding qualified workers – particularly hourly craft workers – to hire. Specifically, 80 percent of firms responding to the survey report they are having a hard time filling craft positions and 57 percent report are having a hard time filling salaried positions. These shortages are prompting many firms to raise wages, improve benefits and expand bonuses and other incentives. Labor shortages are also leading many firms to change the way they operate to become more efficient and less reliant on labor.

The *US Bureau of Labor Statistics* (BLS) reports that construction industry job openings in recent months have been at the highest levels since the series began in December 2000. Meanwhile, the number of unemployed workers with recent construction experience has fallen to record lows. Together, these data show that contractors are having a hard time filling positions and are increasing having to hire workers without construction - or perhaps any - work history. [24]

Another cause for the German labor shortage is the sharp increase in the number of students at universities and technical colleges. By this inflow the number of applicants available on the waged employees market decreases leading thus to an increase of unfilled vacant training places and short-term gaps in the employment system. From a long-term perspective, the latter gaps will be filled by academic graduates. [25] The labour shortage in construction thus embraces in addition to the major occupational groups also the occupational category of civil engineers. Despite the increasing number of academic graduates the number of job vacancies for civil engineers tripled in the period from 2009 to 2018 (Figure 6), and the one for skilled construction workers rose by more than double. Since it is not possible to cover this demand alone by training a high number of unemployed went back to employment. From 2009 to 2018 the number of unemployed civil engineers has fallen by 45 percent and for skilled construction workers by 67 percent. [22]

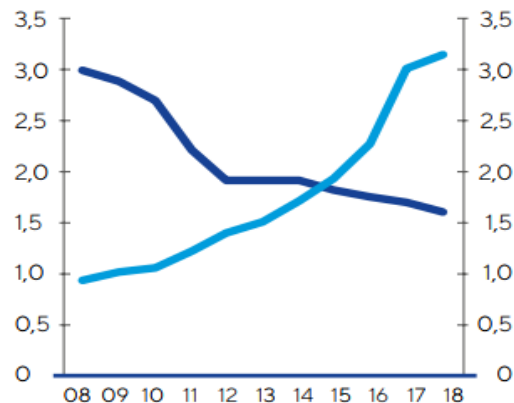

Figure 6. Civil engineers – unemployed (dark blue) and job vacancies (light blue)
(in 1.000) [22]

The risk that qualified staff moves to other sectors seems to be a further concern in the US and likewise in Germany. According to estimates prepared by *SOKA-BAU* [20] employers have developed retention strategies to ensure they are able to avoid losing their apprentices upon completion of their training: While only approximately 5 percent of the apprentices from the 2000 graduation year have been still employed five years later (in 2004) at their training company, this figure increased to 12 percent for the 2008 graduation year. However, alarming is the number of those apprentices leaving the sector right after completion of their training. Only 61 percent of the apprentices from the graduation year 2000 remained on completion in a construction company. Most recently, only around half of these graduates were employed in a construction company right after completion of their training. *SOKA-BAU* reported on the reasons for this sector-exit of skilled workforce by conducting a survey in 2018 with over 900 apprentices, both 300 employers and employees, some instructors from inter-company vocational training facilities and about 230 employees who left the sector. It turned out that around two thirds of the leaving employees are fully trained and skilled and that building companies are more exposed to the exit of employees compared to civil engineering companies. Among the reasons for the exit, the majority cited health reasons, followed by poor economic conditions such as low payment, termination or insolvency of the employer and also a too high workload. Considered as target sectors for new assignments are the manufacturing sector, followed by the public sector and the trade sector. The withdrawal from the construction sector is for around 40 percent of these leaving employees irrevocable and final. However, younger employees (up to the age of 25) were positive about their intentions to return to construction, particularly if the economic conditions in the construction market change.

A number of conclusions can be drawn from the survey. On the one hand the surveyed persons confirm that the exit of skilled workforce presents a pressing problem which will become even more important in the coming years. On the other hand is obvious, that the hard physical work is still considered a special challenge for the employees. This supplements the data for the pension entry of employees available to *SOKA-BAU* [20]. Thereafter, significantly more new retirees in construction received a pension due to partial invalidity or full invalidity relative to new pensioners from other sectors.

The demographic shifts have made the issue of ensuring healthier workers, especially those of advanced age, much more pressing. Because of the physical demands of the work, construction workers who are employed have to be healthier than the general population, but the same physical demands cause workers with injuries or illness to leave the industry. To address this issue, research and policies are urgently needed to identify and promote effective programs and intervention techniques and strategies that can meet the safety and health needs of older workers.

Age-equitable and age-appropriate workplaces are also needed, not only to reduce the incidences of early retirement (because of a reduction in earning capacity), but also in order to be able to offer young skilled workers a long-term perspective and with a strong focus on retaining skills in the sector. This requires opportunities for personal development and career promotion and also the facilitating of horizontal career routes, which is aimed at developing a work structure suited to the elderly. This also includes to make better use of the possibilities for technical progress, whereby the construction sector invests traditionally and comparatively very little in equipment.

A majority of companies has recognized the problem of shrinking supply of skilled manpower, and has therefore implemented numerous measures. The training of young specialists is the preferred instrument including the spectrum of dual courses of study and training opportunities with supplementary qualification. In addition, the focus lies on different instruments of staff retention and the continual planning of human resources requirements. [25] Innovative methods of recruiting and the enticement of employees by competitors is becoming increasingly important. Meanwhile approximately 12 percent of the companies recruit employees that way, compared to 1 percent ten years back. [26] Further channels for recruitment are e.g. the setup of candidate pools by job boards, the transregional search or the addressing of suitable candidates with future potential. [27]

## 4. CONCLUSION AND OUTLOOK

This paper discusses the subject of qualitative (eg management quality) and quantitative (eg workforce shortage) aspects relating to human resources in construction and their impact on productivity.

The image that people have of the construction industry as an employer is a relatively poor one, with inadequate gender diversity and little job security (owing to the cyclical nature of the business). As a result, construction companies often struggle to attract talented recruits to their workforce. Relative to companies in other industries, construction companies engage less often and less effectively in internal people-development initiatives. [28]

There is substantial heterogeneity in management practices across establishments, with small firms having lower scores than large firms on average with a robust positive and economically important association between the management score and establishment level productivity. This association increases with firm size. Comparison of similar surveys in the US and Germany indicates that the average management score is lower in Germany than in the US. Overall, the results point towards lower management scores being at least in part to blame for the differences in aggregate productivity between Germany and the US.

Unquestioned opinion of representatives and experts suggests that qualification of employees and labor (workers) is a crucial competitive factor for the construction industry. The quality of training for skilled labor, the advanced training for the middle management (site managers and foremen/supervisors linked to the job site) and the university/college education of architects and civil engineers is therefore given high attention by the construction sector. The high professional competences of civil engineers which also contribute to the international competitiveness of the construction business, and also the input of highly qualified workers of all levels are the fundamental basis for the realization of the typical form of incremental innovation in construction. New techniques, new materials and advanced construction equipment can be integrated in the production smoothly and without any major effort, based on a high level of qualification and process expertise of the employees concerned and likewise without further development of their competences in the use of new equipment and the application of new or advanced materials.

Further, recruitment problems represent a limiting factor for the construction industry. The bottleneck of skilled labor has become the biggest business risk in construction. In 2019 the industry-wide job vacancies clearly outstripped labor supply in German construction. German industrial sectors are not affected by the current skilled-labor shortage to the same degree as companies in construction though the increase was even greater over the last ten years however not the absolute volume relative to construction. [22]

The skilled-labor shortage results in various consequences and outcomes. On the one hand the durations to fill job vacancies increases and on the other hand the negotiating position of employees towards employers improves by the mismatch between supply and demand. The strengthened and unbalanced position leads to above-average wage and salary progressions for enterprises. [29] In addition, companies are reaching their capacity limits (in times of a sound construction sector) because of the lack of available skilled workforce or because this is not recognized in a timely manner (Figure 7). Apart from sales losses competitiveness and productivity decline. Thus, vacancies can result in potential economic consequences for the enterprises in construction. [27]

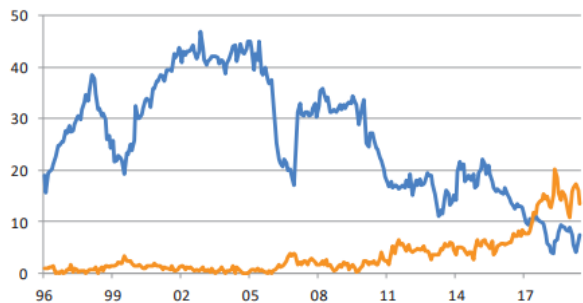

Figure 7. Disruption through lack of orders (dark blue) and shortage of labour (light orange)
in the main construction sector (in percent) [20]

Many variables have an impact on how the labor shortage will develop in the future. Migration, employment and educational behavior or the demographic change have an impact on this development. Forward-looking prognoses by experts are regularly published. This allows statements to be made regarding the development of labor shortage. The investigations, however, show that the verification by means of figures and indicators is difficult and that empirical or statistical forecasts do not reflect the subjective perception of construction companies concerned. Further, the differing views reveal how difficult it is to predict future demand and supply for workforce and to provide reliable statements concerning an existing or future qualitative or quantitative lack. It is positive to note that the sector identified and implemented countermeasures at an early stage. However, it remains to be seen, whether these approaches will have the desired effect.

Construction is likely to be affected to a larger extent by the lack of digital (IT) specialists, because the construction sector enjoys a rather modest reputation and appeal as an employer for this target group. This may only change by an extensive sector wide modernization. In this regard the primary responsibility lies with the higher education sector and also the private sector to develop digital competencies. [30, 31] The construction industry in Germany records a relatively low degree of investments in ICT respectively digitalization, also against the background of the industry's low productivity growth as compared to other sectors.

Recent technological progress in design, manufacturing, and logistics is already inspiring new companies to introduce more efficient ways of building. If they succeed, the construction industry could ultimately go the way of agriculture and manufacturing (and retail), with the associated mix of societal gain and upheaval. Consumers will reap the benefits of an industry that is more productive despite employing fewer workers, but many of those workers will be aptly-skilled newcomers who supplant veteran workers that struggle to adapt.

Despite the proven ability of new technologies to lift productivity in other industries, construction lags significantly behind other sectors in its use of innovative digital tools, and is slow to adopt new materials, methods, and technology. The objective of further research is to develop a framework - pertinent to the German industry - that can aid decisionmakers in determining how innovation could be employed to increase project productivity, given the qualitative (eg management practices/quality and mismatch of skilled labour) and quantitative asdpects (eg skilled-labour shortage).

**AUTHORS**

**Professor Dr. Matthias Bahr** Dean Department of Civil Engineering Hochschule Biberach, Germany.

**Mr. Leif Laszig** Master in Civil Engineering, Master in Economics doctoral student with more than 20 years work experience.